\documentclass[aps,pra,twocolumn,showpacs,nofootinbib]{revtex4-1}
\usepackage {amsmath}
\usepackage {amssymb}
\usepackage[dvips]{graphicx}
\usepackage[english]{babel}
\usepackage{indentfirst}

\usepackage[usenames]{color}
\usepackage{ulem}
\usepackage[colorlinks=true]{hyperref}

\renewcommand{\vec}[1]{\boldsymbol{#1}}

\newcommand{\rmi}{\mathrm{i}}
\newcommand{\rme}{\mathrm{e}}

\newcommand{\p}{\partial}

\begin{document}

\title{Out-of-phase few-cycle solitons in multicore fibers}

\author{S.\,A.\,Skobelev}
\author{A.\,A.\,Balakin}
\author{E.\,A.\,Anashkina}
\author{A.\,V.\,Andrianov}
\author{A.\,G.\,Litvak}
\email{sksa@ufp.appl.sci-nnov.ru}

\affiliation{Institute of Applied Physics RAS, 603950 Nizhny Novgorod, Russia}

\date{\today}

\begin{abstract}
	An equation is derived for analyzing the self-action of a wave packets with few optical cycles in multicore fibers (MCF). A new class of stable out-of-phase spatio-temporal solitons with few cycle durations in the MCF with cores located in a ring is found and analyzed. The stability boundary of the obtained solutions is determined. As an example of using such solitons, we considered the problem of their self-compression in the process of multisoliton dynamics in the MCF. The formation of laser pulses with a duration of few optical cycles at the output of a ten-core MCF is shown.
\end{abstract}

\maketitle
\section{Introduction}\label{sec:1}

In the last decade, significant progress has been made in the generation of high-energy laser pulses with a small number of field oscillations. These advances in laser science have led to the development of a new field of \textit{extreme light}, in which the interaction of such extremely short pulses with matter is studied. A stable propagation mode of such few cycle pulses is realized in waveguide systems with anomalous group velocity dispersion and is associated with the existence of exact solutions of the wave equation in the form of solitons \cite{Skobelev_2007, Sakovich_2006, Skobelev_2011, Leblond_2013}.

The concept of optical solitons has played an important role in the development of nonlinear optics in recent years. Several remarkable applications of the use of solitons should be noted: supercontinuum generation \cite{Agrawal_2019, Dai_2018, Petersen_2014, Martinez_2018}; self-compression of laser pulses to a duration of a small number of field oscillations \cite{Agrawal_2019, Skobelev_2007, Nazarkin_1999, Mollenauer_1983, Li_2010,Anashkina_2011}; creation of effective sources of laser pulses in insufficiently developed spectral ranges, such as mid-IR \cite{Agrawal_2019, Anashkina_2020, Duval_2016, Tang_2016} and ultraviolet \cite{Russell_2011, Skobelev_2015, Biancalana_2011, Russell_2013} through the use of various types of non-stationary nonlinearities (Raman and ionization ones).

Along with this, the search for stable non-one-dimensional soliton solutions is of interest. However, it is well known that in a bulk medium with cubic nonlinearity, the wave field is subject to filamentation instability and self-focusing \cite{Bespalov_1966, Zakharov_1974, Boyd_2009}. In recent decades, theoretical and experimental studies of the nonlinear dynamics of wave fields in multi-core fibers (MCFs), consisting of identical equidistantly spaced weakly coupled cores, have intensified in order to eliminate filamentation instability in the transverse direction, which is a fundamental limitation on the generation and use of high-power laser pulses. A number of interesting results have been obtained in such MCFs: supercontinuum generation \cite{Tran2014,Babushkin07, Tran2014a}, the laser pulse compression \cite{Aceves,Rubenchik2015,Skobelev2019,Balakin2019,Tran2014,Tran2014a,Cheskis03, Andrianov_2019, Rubenchik2015a}, nonlinear switching \cite{Andrianov2019} and the light bullet formation \cite{Tran2014,Tran2014a,Cheskis03,Eilenberger2011,Balakin2019,Lobanov2010,Minardi2010,Leblond2017,Rubenchik2015,Rubenchik2015a, Andrianov_2019}. However, most of them were obtained for in-phase field distributions, which are influenced by a discrete analogue of filamentation instability and stochasticity  \cite{Skobelev_2018}.

At the same time, out-of-phase distributions of a wave field in MCFs of various configurations, which are not subject to filamentation instability, were found \cite{Tunnermann_2015, Balakin_2018, Skobelev_2020, Skobelev_2021, Balakin_2020}. The existence of out-of-phase modes make it possible to find stable out-of-phase spatio-temporal solitons in MCFs within the slowly varying amplitude approximation \cite{Skobelev2019, Balakin_2020, Litvak_2020}. Such spatio-temporal solitons allow to operate with significantly higher total energy in comparison with single-core fibers. The question arises of the existence of similar solutions describing stable spatio-temporal soliton-like structures with an \textit{arbitrary number} of field oscillations. Their presence will make it possible to generalize the well-known methods of compression of laser pulses in a single fiber as applied to MCFs and to take a significant step towards solving the problem of creating laser pulses of high energy and short duration in systems built entirely in fiber format.

In this work, we derive the \emph{basic} equation describing the evolution of the wave field in MCF without scale separation into slow envelope and high-frequency carrier (Section \ref{sec:2}). We find and analyze a new class of stable solitary solutions of the wave field, which describes the propagation in a transparent nonresonant media of out-of-phase spatio-temporal soliton-like structures of circularly polarized optical pulses, including a few-cycle ones (Section \ref{sec:3}). We analyze the stability of the found solutions and demonstrate their relation to the spatio-temporal solutions \cite{Skobelev2019} found in the framework of the nonlinear Schr\"odinger equation (Section \ref{sec:4}). We generalize the concept of high-order Schr\"odinger solitons as applied to the found spatio-temporal solitary solutions with a small number of field oscillations and demonstrate its use for effective compression of laser pulses in an MCF (Section~\ref{sec:5}).

\section{Basic equations}\label{sec:2}

To adequately describe the space-time evolution of ultrashort circularly polarized laser pulses ($\mathcal{E}=\mathcal{E}_x+\rmi\mathcal{E}_y$, where $\mathcal{E}_x$ and $\mathcal{E}_y$ are the corresponding components of the electric field strength) in MCF with cubic nonlinearity, we turn directly to the wave equation
\begin{equation}\label{eq:1}
	\dfrac{\partial^2\mathcal{E}}{\partial z^2}+\Delta_{_\perp}\mathcal{E}-\dfrac{1}{c^2}\dfrac{\partial ^2}{\partial t^2}\int_{-\infty}^{t}\varepsilon(x,y,t-t')\mathcal{E}(t')dt'=\dfrac{4\pi}{c^2}\dfrac{\partial^2\mathcal{P}_{nl}}{\partial t^2} .
\end{equation}
Here $\mathcal{P}_{nl}$ is the nonlinear response of the medium, $c$ is the speed of light in vacuum. The linear dielectric constant of the medium $\varepsilon$ can be approximated as \cite{Skobelev_2007}
\begin{equation}\label{eq:3}
\varepsilon(\omega) \approx \varepsilon_0-\frac{\omega_D^2}{\omega^2}+b\omega^2,
\end{equation}
where $\varepsilon_0 $ is the static dielectric constant, $\omega_D^2 $ and $b$ characterize the low-frequency and high-frequency dispersion.

Note that the dependence \eqref{eq:3} describes the dispersion of transparent dielectrics with a high degree of accuracy in the near IR and part of the visible spectral range. For example, the dielectric constant of fused silica can be written through the Sellmeyer formula \cite{Marcuse}
\begin{equation}\label{eq:4}
	\varepsilon_\text{glass}= 1+\frac{B_1}{1-A_1/\lambda^2}+\frac{B_2}{1-A_2/\lambda^2}+\frac{B_3}{1-A_3/\lambda^2}  ,
\end{equation}
where $B_1=0.696$, $B_2=0.4079$, $B_3=0.897$, $A_1=4.62\times10^{-3}~\mu \text{m}^2$, $A_2=1.36\times10^{-2}~\mu \text{m}^2$, $A_3=98~\mu \text{m}^2$. Out of resonances $\lambda_2 < \lambda < \lambda_1$ (where $\lambda_1=9.896~\mu\text{m}=2\pi c/\omega_1$ and $\lambda_2=0.116~ \mu\text{m}=2\pi c/\omega_2$) the expression \eqref{eq:4} can be rewritten in a form similar to the expression \eqref{eq:3}
\begin{equation}\label{eq:5}
	\varepsilon_\text{glass}\approx (1+B_1+B_2)+\frac{1}{\lambda^2}\left(B_1A_1+B_2A_2\right)-\frac{B_3}{A_3}\lambda^2 .
\end{equation}

\subsection{Linear case}

Next, let us analyze the propagation of laser radiation in an array of parallel weakly coupled cores without taking into account the nonlinearity of the medium $\mathcal{P}_\text{nl}(\mathcal{E})=0$. As an example, the perturbation of the dielectric constant of a medium can be considered in the form $\delta \varepsilon(x,y) =\delta\varepsilon_0 \sum_n \exp\left[-\tfrac{((x-x_n)^2+(y-y_n)^2)^2}{r_n^4}\right],$ where $x_n$, $y_n$, $r_n$ are the position and radius of the cores, $\delta\varepsilon_0 \ll \varepsilon_0$ is the difference in permittivity of cores and cladding. Obviously, perturbations of the dielectric constant of the medium will lead to the dependence of the coefficients $\varepsilon_0$, $\omega_D$, $b$ on the transverse coordinates $x, y$.

Substitution of the expression \eqref{eq:3} into the equation \eqref{eq:1} gives the wave equation
\begin{multline}\label{eq:7}
	\frac{\partial^2\mathcal{E}}{\partial z^2}+\Delta_{\perp}\mathcal{E} - \frac{ \varepsilon_0(x,y)}{c^2}\frac{\partial^2\mathcal{E}}{\partial t^2} - \\ - \frac{ \omega_D^2(x,y)}{c^2} \mathcal{E}+\dfrac{ b(x,y)}{c^2}\frac{\partial^4\mathcal{E}}{\partial t^4} =0,
\end{multline}
having action
\begin{multline}\label{eq:8}
	S = \iiiint \bigg[\dfrac{ \varepsilon_0(x,y)}{c^2} \left|\dfrac{\partial\mathcal{E}}{\partial t}\right|^2 - \dfrac{ \omega_D^2(x,y)}{c^2} |\mathcal{E}|^2+\\+
	\dfrac{ b(x,y)}{c^2}\left|\dfrac{\partial^2\mathcal{E}}{\partial t^2}\right|^2 - \left|\dfrac{\partial\mathcal{E}}{\partial z}\right|^2 - |\nabla_\perp \mathcal{E}|^2 \bigg]dxdydzdt .
\end{multline}

We use the approximation that the fundamental guided modes of optical waveguides weakly coupled to each other \cite{Agrawal_2019}. In this case, the propagation of laser radiation in the MCF can be approximately described as a superposition of modes localized in each cores
\begin{equation}\label{eq:9}
	\mathcal{E}(z,t,x,y) \approx \sum\limits_{n}\mathcal{A}_n(z,t)\phi(\vec{r}_\perp - \vec{r}_n),
\end{equation}
where $\phi(\vec{r})$ is the transverse structure of the fundamental mode, $\mathcal{A}_n$ is the field envelope in the $n$-th cores.

Assuming the overlap of wave fields from neighboring cores to be small $\int \phi(\vec{r}) \phi(\vec{r}) dxdy \gg \int \phi(\vec{r}) \phi(\vec{r}+\vec{d}) dxdy$, where $\vec{d}$ is the lattice period, we can apply the variational approximation, i.e. substitute the field in the form of the sum \eqref{eq:9} into the expression \eqref{eq:8} and integrate over the transverse coordinates (below $\phi=\phi(\vec{r})$, $\phi_+=\phi(\vec{r}+\vec{d})$)
\begin{widetext}
\begin{multline}\label{eq:10}
	S = \sum_n \iint  \Bigg[ \left|\frac{\partial \mathcal{A}_n}{\partial t}\right|^2 \underbrace{\iint \dfrac{ \varepsilon_0(x,y)}{c^2} \phi^2 dxdy}_{\beta} + \left(\frac{\partial \mathcal{A}_n^*}{\partial t} \frac{\partial \mathcal{A}_{n+1}}{\partial t} +c.c.\right) \underbrace{\iint \dfrac{ \varepsilon_0(x,y)}{c^2} \phi \phi_+ dxdy}_{\beta_1} - \\ -
	|\mathcal{A}_n|^2 \underbrace{\iint \left(\dfrac{ \omega_D^2(x,y)}{c^2} \phi^2 + (\nabla_\perp \phi)^2\right) dxdy}_{\sigma} + (\mathcal{A}_n^* \mathcal{A}_{n+1}+c.c.) \underbrace{\iint \left(\dfrac{ \omega_D^2(x,y)}{c^2} \phi \phi_+ + \nabla_\perp \phi \nabla_\perp \phi_+\right) dxdy}_{-X} + \\ +
	\left|\frac{\partial^2 \mathcal{A}_n}{\partial t^2}\right|^2 \underbrace{\iint \dfrac{ b(x,y)}{c^2} \phi^2 dxdy}_{\gamma} + \left(\frac{\partial^2 \mathcal{A}_n^*}{\partial t^2} \frac{\partial^2 \mathcal{A}_{n+1}}{\partial t^2} +c.c.\right) \underbrace{\iint \dfrac{ b(x,y)}{c^2} \phi \phi_+ dxdy}_{\gamma_1} - \\ -
	\left|\frac{\partial \mathcal{A}_n}{\partial z}\right|^2 \underbrace{\iint \phi^2 dxdy}_{\alpha} - \left(\frac{\partial \mathcal{A}_n^*}{\partial z} \frac{\partial \mathcal{A}_{n+1}}{\partial z} +c.c.\right) \underbrace{\iint \phi \phi_+ dxdy}_{\alpha_1}
	\Bigg]dtdz .
\end{multline}
\end{widetext}
Note that quantities $\varepsilon_0$, $\omega_D$ and $b$ can be considered almost constant on the scale of the cores for the fibers under consideration. Therefore, we can put
\begin{multline}
	\alpha_1 \approx \zeta \alpha,  \quad \beta_1 \approx \zeta \beta, \\ \gamma_1 \approx \zeta \gamma , \quad \zeta = \frac{\iint \phi \phi_+ dxdy} {\iint \phi^2 dxdy} \ll 1 .
\end{multline}
In this case, the action \eqref{eq:10} generates the equations
\begin{multline}\label{eq:10_eq}
	\left(\beta \frac{\partial^2}{\partial t^2} - \alpha \frac{\partial^2}{\partial z^2} - \gamma \frac{\partial^4}{\partial t^4} \right)\left[\frac{}{} \mathcal{A}_n + \zeta \mathcal{A}_{n-1} + \zeta \mathcal{A}_{n+1}\right] + \\
	+\sigma \mathcal{A}_n - X(\mathcal{A}_{n-1} + \mathcal{A}_{n+1})= 0.
\end{multline}

For a further analysis, it is convenient to use the evolutionary equation for the field in the simplest form of the reduced wave equation. Assuming the changes in the field distributions $\mathcal{A}_n(z,t)$ to be small on scales of order of wavelengths, and neglecting the reflected wave, we obtain for a wave field traveling along the $z$ axis
\begin{multline}\label{eq:11_eq}
	\frac{\partial^2 \mathcal{A}_n}{\partial t^2} - \frac{\alpha}{\beta} \frac{\partial^2 \mathcal{A}_n}{\partial z^2} = \left[\frac{\partial}{\partial t} - V \frac{\partial}{\partial z}\right] \left[\frac{\partial}{\partial t} + V \frac{\partial}{\partial z}\right] \mathcal{A}_n \approx\\\approx 2 V \frac{\partial^2 \mathcal{A}_n}{\partial z \partial \tau} ,
\end{multline}
where $V=\sqrt{{\alpha}/{\beta}}$ is the speed of light in the medium, $\tau = t-z/V$ is the time in the accompanying coordinate system. The formal condition for the applicability of the approximation \eqref{eq:11_eq} is that the envelope changes slowly along the propagation path: $|\partial_z \mathcal{A}_n| \ll |\partial_\tau \mathcal{A}_n|/V$. In the case of quasi-monochromatic radiation, this approach leads to an Schr\"odinger-like equation for the envelope. Application the approximation \eqref{eq:11_eq} to Eq.~\eqref{eq:10_eq} gives
\begin{multline}\label{eq:10_eq1}
	2\sqrt{\alpha\beta}\frac{\partial^2}{\partial z \partial \tau} \left(\frac{}{}\mathcal{A}_n + \zeta \mathcal{A}_{n-1} + \zeta \mathcal{A}_{n+1}\right)+ \\+
	\sigma \mathcal{A}_n - X(\mathcal{A}_{n-1} + \mathcal{A}_{n+1}) -\\- \gamma \frac{\partial^4}{\partial \tau^4} \left(\frac{}{}\mathcal{A}_n + \zeta \mathcal{A}_{n-1} + \zeta \mathcal{A}_{n+1}\right) = 0.
\end{multline}

The terms with $\zeta \ll 1$ describe the weak influence of the field of neighboring cores on the velocity $V$ and on the high-frequency dispersion of the medium $\gamma$. Variable change $\mathcal{A}_n^\text{new} = \mathcal{A}_n + \zeta \mathcal{A}_{n+1} + \zeta \mathcal{A}_{n-1}$ allows one to get rid of them, in contrast to the much stronger influence of the terms with $X$, which describe the field coupling in neighboring cores. As a result, we obtain, up to $\zeta^2$ terms, a unidirectional wave equation in new variables (we omit the ``new'' subscript) that describe the dynamics of a laser pulse in an MCF without taking into account the nonlinearity
\begin{equation}\label{eq:13}
	2 \sqrt{\alpha\beta} \frac{\p^2 \mathcal{A}_n}{\p z \p \tau} + \sigma \mathcal{A}_n - (X+\sigma \zeta) (\mathcal{A}_{n+1}+\mathcal{A}_{n-1}) - \gamma \frac{\p^4 \mathcal{A}_n}{\p \tau^4}=0 .
\end{equation}

It is seen from the expression \eqref{eq:10} that the distribution of the fundamental mode $\phi(x,y)$ contributes to the ``low-frequency'' dispersion $\sigma$ of the combined medium. Moreover, this contribution can be significant in the case of cores with a small radius
\begin{equation}
	\iint\frac{ \omega_D^2(x,y)}{c^2} \phi^2dxdy< \iint(\nabla_\perp \phi)^2dxdy .
\end{equation}
Moreover, the terms $\sigma $ and $X$ can be of the same order due to the different structure of the integrands. The presence of the gradient $\nabla_\perp \phi \nabla_\perp \phi_+$ in the integrand for $X$ leads to the fact that the corresponding term for weakly coupled cores becomes negative
\begin{multline}
X+\sigma \zeta \approx \frac{\iint \phi \phi_+ dxdy}{\iint \phi^2 dxdy} {\iint (\nabla_\perp \phi)^2 dxdy} -\\- \iint  \nabla_\perp \phi \nabla_\perp \phi_+ dxdy >0 .
\end{multline}

\subsection{Nonlinear case}

Let us take into account the influence of the media nonlinearity $\mathcal{P}_\text{nl}$ on the dynamics of the laser pulse. The term $\mathcal{P}_\text{nl}$ in the equation \eqref{eq:1} takes into account the nonstationarity of the nonlinear response of the medium
\begin{multline}\label{eq:14}
	\mathcal{P}_\text{nl, n}=n_2(1-f_R)|\mathcal{A}_n|^2\mathcal{A}_n+\\+n_2f_R\mathcal{A}_n\int\limits_0^\infty|\mathcal{A}_n(\tau-\tau')|^2h_R(t')d\tau' ,
\end{multline}
where $ f_R $ is the partial contribution of the inertial SRS response to nonlinear polarization, $ n_2 $ is the nonlinearity coefficient. Here we have assumed that the nonlinearity is the same in all cores. The Raman response function $ h_R $ is responsible for the Raman gain and can be determined from the experimentally measured Raman spectrum. An approximate analytical form of this function for silica fibers is as follows \cite{Agrawal_2019}
\begin{equation}\label{eq:15}
	h_R(\tau)=\dfrac{\tau_1^2+\tau_2^2}{\tau_1\tau_2^2}\exp(-\tau/\tau_2)\sin(\tau/\tau_1),
\end{equation}
where $\tau_1=12.2$ fs, $\tau_2=32$ fs, $f_R=0.18$ for silica fiber.

As a result, we arrive at the following final unidirectional wave equation in dimensionless form describing the self-action of a laser pulse with an arbitrary duration in the MCF
\begin{multline}\label{eq:16}
	\frac{\p^2 u_n}{\p \hat{z} \p \hat{\tau}}+u_n-\chi(u_{n-1}+u_{n+1})-\mu \frac{\p^4u_n }{\p\hat{\tau}^4}+\\+
	\frac{\p^2}{\p\hat{\tau}^2}\bigg[(1-f_R)|u_n|^2 u_n+\\+ f_R u_n\int_0^\infty|u_n(\hat{\tau}-\hat{\tau}')|^2 \hat{h}_R(\hat{\tau}')d\hat{\tau}'\bigg]=0 .
\end{multline}
Here we introduced normalizing dimensional factors for time $\tau_0=1/\omega_0$, traces $z_0=2\sqrt{\alpha\beta}\omega_0/\sigma$, fields $\mathcal{A}_0=\sqrt{\sigma c^2/(4\pi n_2 \omega_0^2)}$, $\mu=\gamma\omega_0^4/\sigma$, $\chi=\zeta + X/\sigma$, $\hat{h}_R = h_R(\tfrac{\hat{\tau}'}{\omega_0})/\omega_0$, $\omega_0$ - carrier frequency. In what follows, we will not write the hat sign.

In the case under consideration, the linear dispersion law for waves propagating along $z$, in the case of uncoupled cores ($\chi=0$), has the form
\begin{equation}
	k=-\frac1{\omega}+\mu\omega^3 .
\end{equation}
Note that by changing the center frequency $\varpi$ of broadband radiation, one can control the role of dispersion in the dynamics of the system. In particular, for radiation with a frequency $\omega_\text{bnd} = 1/\left(3\mu\right)^{1/4}$, the group velocity dispersion parameter
\begin{equation}\label{eq:beta2}
	\beta_2=\frac{\partial^2k}{\partial\omega^2}=-\frac{2}{\omega^3}+6\mu\omega
\end{equation}
become zero. Accordingly, for wave fields with a frequency of $\varpi\gg\omega_\text{bnd}$, the field spectrum is concentrated in the region with normal group velocity dispersion, and for $\varpi\ll\omega_\text{bnd}$, the dispersion is anomalous.

It is important to note that the obtained equation \eqref{eq:16} has a fundamental character and allows one to describe the nonlinear dynamics of arbitrarily short laser pulses in weakly coupled multicore fibers. It is easily generalized to the case of MCF with a core configuration other than circular or linear too.

\section{Out-of-phase spatio-temporal few-cycle soliton}\label{sec:3}

This section is devoted to the analysis of the most interesting case when the main role in the dynamics of a laser pulse is played by the low-frequency dispersion of the medium ($\mu = 0$) and the media nonlinearity being inertialess ($f_R = 0$).

Consider an MCF in which the cores are arranged in a ring. Earlier, we found a number of stable nonlinear solutions for wave beams propagating in the considered MCF \cite{Balakin_2018, Skobelev_2020, Skobelev_2021, Balakin_2020}. The most interesting of them is the $\pm$-mode $u_n \propto (-1)^n$, which provides coherent transportation of maximal power at a given field amplitude. Moreover, this solution is stable and exists at all amplitudes.

Along with this, soliton solutions are of interest. We found out-of-phase soliton solutions in the framework of the NSE equation \cite{Skobelev2019}
\begin{equation}\label{eq:17}
	u_n=(-1)^n\frac{\sqrt{2}a_0\rme^{\rmi (2\chi-a_0^2)z}}{\cosh(a_0\tau)} .
\end{equation}
It was shown that the solution \eqref{eq:17} is stable with respect to filamentation instability. The analysis was carried out using the {\it 2nd Lyapunov method}.

An interesting question is the existence of a stable out-of-phase spatio-temporal soliton with a \textit{small number of field oscillations} in the MCF within the framework of the derived equation \eqref{eq:16}. The found solution will provide coherent propagation of few-cycle laser pulses of constant shape in all available MCF cores. In this case, the total energy of the found nonlinear structure will significantly exceeds the soliton energy in a single-core fiber.

In the case of MCF, in which the cores are located on a ring, the amplitude of the nonlinear structure is the same in all cores. So, we look for a solution in the form
\begin{equation}\label{eq:18}
	u_n(z,\tau)=(-1)^n {u}_\pm(z,\tau) .
\end{equation}
Substituting the expression \eqref{eq:18} into \eqref{eq:16} and neglecting the nonlinearity inertia ($f_R = 0$), we obtain the equation for the wave field dynamics in MCF
\begin{equation}\label{eq:19}
	\frac{\p^2 {u}_\pm}{\p z \p\tau}+(1+2\chi){u}_\pm+\frac{\p^2 }{\p\tau^2}\left(|{u}_\pm|^2{u}_\pm \right)=0 .
\end{equation}
Note that the absolute value of the coefficient related to the dispersion properties of the MCF in the out-of-phase mode is $(1+2\chi)$ times greater than for a single-core fiber, which agrees with the results of paper \cite{Skryabin_2009}.

It follows from Eq.~\eqref{eq:19} the following integral relation for localized field distributions
\begin{equation}\label{eq:20}
	\int_{-\infty}^{+\infty} u_\pm d\tau= 0 ,
\end{equation}
expressing the absence of a averaged field in such distributions and indicating their oscillating character. The equation \eqref{eq:19} belongs to the class of Hamiltonian systems with a Hamiltonian of the form
\begin{equation}\label{eq:21}
	H_\pm=\int_{-\infty}^{+\infty}\left[\frac12|{u}_\pm|^4-(1+2\chi)\left|\int_{-\infty}^{\tau}{u}_\pm d\eta \right|^2 \right]d\tau .
\end{equation}

Earlier we found and analyzed a new class of stable soliton solutions in the framework of the equation \eqref{eq:19} \cite{Skobelev_2007}. The wave solitons of the equation can be represented by a two-parameter family of solutions of the form
\begin{equation}\label{eq:22}
	{u}_\pm(z,\tau)=\sqrt{v_s}G(\xi)\exp[i\omega_s(\tau+v_s z)+i\varphi(\xi)] ,
\end{equation}
where $\omega_s$ is the characteristic carrier frequency, $v_s$ is the parameter that determines the group velocity of the soliton, $\xi = \omega_s (\tau-v_s z) $. The phase and group velocities are different, which leads to oscillations of the wave structure, which will be noticeable for short laser pulses. The envelope of the soliton $G(\xi)$ and the nonlinear phase $\varphi(\xi)$ obey the following equations:
\begin{subequations}\label{eq:23}
\begin{gather}
	\dfrac{d\varphi}{d\xi}=\dfrac{G^2(3-2G^2)}{2(1-G^2)^2} , \label{eq:23a}\\
	\int_{G_m}^G\dfrac{1-3G^2}{G\sqrt{\delta^2-F(G^2)}}dG=\pm(\xi-\xi_0) , \label{eq:23b}
\end{gather}
\end{subequations}
where $F(G^2)=G^2\left[3/2(1+\delta^2)-(4-5G^2)/4(1-G^2)^2\right]$,
$G_m$ is the maximum amplitude of the soliton, $\xi_0$ is the integration constant corresponding to the position of the maximum of the envelope of the soliton. As seen from \eqref{eq:23b}, the solutions for the envelope of the soliton $G(\xi)$ depend only on the parameter $\delta^2 = (1+2\chi)/(\omega_s^2 v_s)-1$ and exist at $0 \leq \delta \leq \delta_\text{cr} \equiv \sqrt{1/8}$. An important feature of the considered wave solitons is the semi-bounded spectrum of their admissible solutions, i.e., the presence of a boundary solution corresponding to the limiting soliton with the minimum possible pulse duration and, accordingly, with the maximum possible amplitude. It should be noted that the existence of a limiting soliton with the shortest duration is determined by the integral \eqref{eq:20}. The duration of the \textit{limiting soliton} for $\delta =\delta_\text{cr}$ is $\tau_s^* \approx 2.31\omega_s^{-1}$.

Thus, the out-of-phase solitary solution of the wave field $u_n$ with an arbitrary number of field oscillations in an MCF consisting of cores arranged in a ring, with an inertialess nonlinearity of the Kerr type, has the form
\begin{multline}\label{eq:24}
	u_n^\text{sol}(z,\tau)=(-1)^n\sqrt{v_s}G(\xi)\exp[i\omega_s(\tau+v_s z)+i\varphi(\xi)], \\ \delta^2=\frac{1+2\chi}{\omega_s^2 v_s}-1\leq  \frac18.
\end{multline}

\begin{figure}[tpb]
\centering
\includegraphics[width = \linewidth]{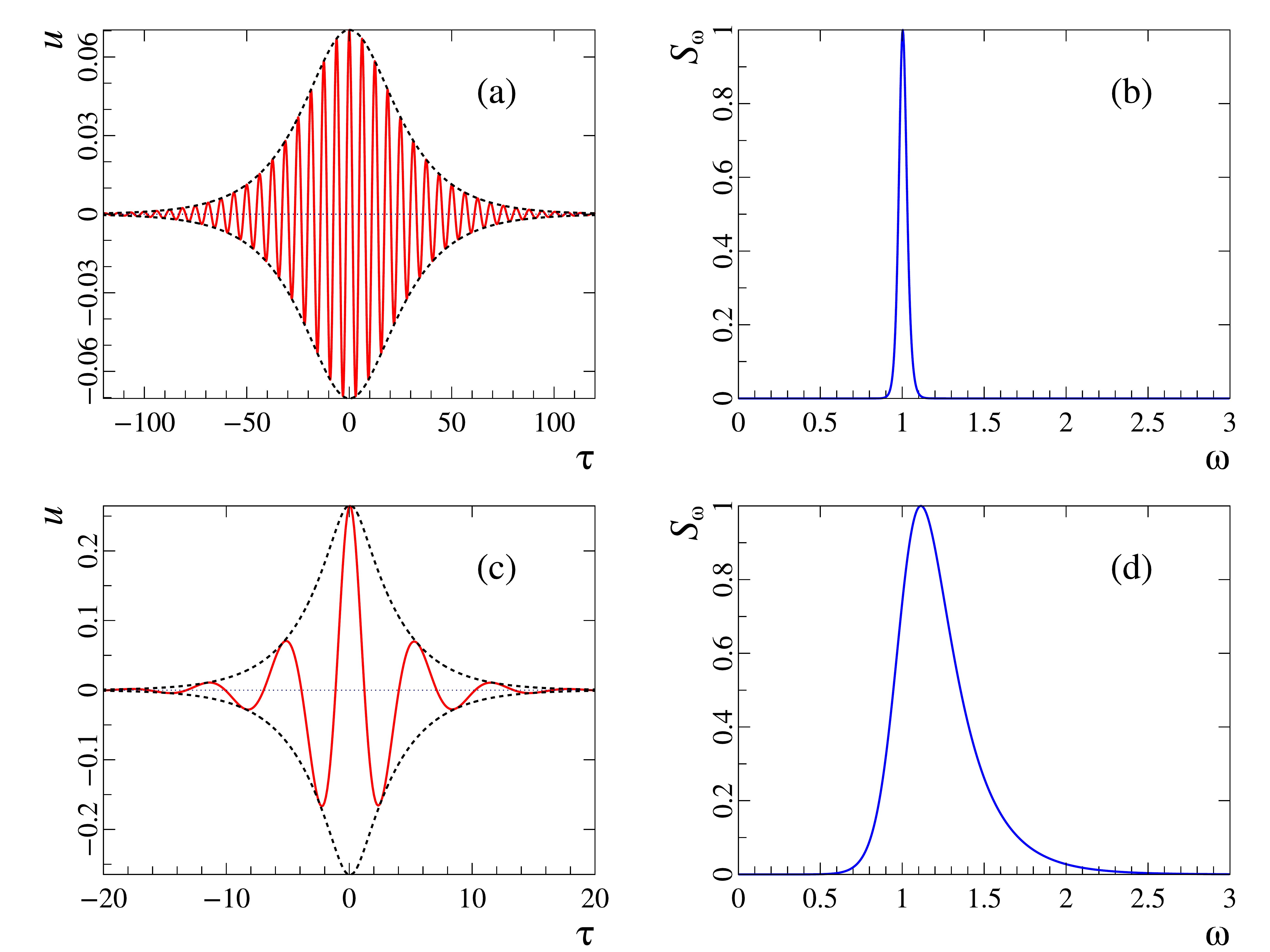}
\caption{
	Exact soliton solutions for one of the field polarizations (red solid curves) corresponding to $\delta = 0.06$, $\omega_s = 1$ {\bf (a)} and $\delta = 0.3$ {\bf (c)}. The dotted black line represents the distribution of the field envelope $\sqrt{v_s} G(\xi)$. Figures {\bf (b)} and {\bf (d)} show the spectral intensity distributions for different $\delta$: {\bf (b)} for $\delta = 0.06$; {\bf (d)} for $\delta = 0.3 $. Coupling coefficient is $\chi = 0.3$, the central frequency $\omega_s = 1$.}\label{ris:ris1}
\end{figure}

\begin{figure*}[tp]
\centering
\includegraphics[width =0.8 \linewidth]{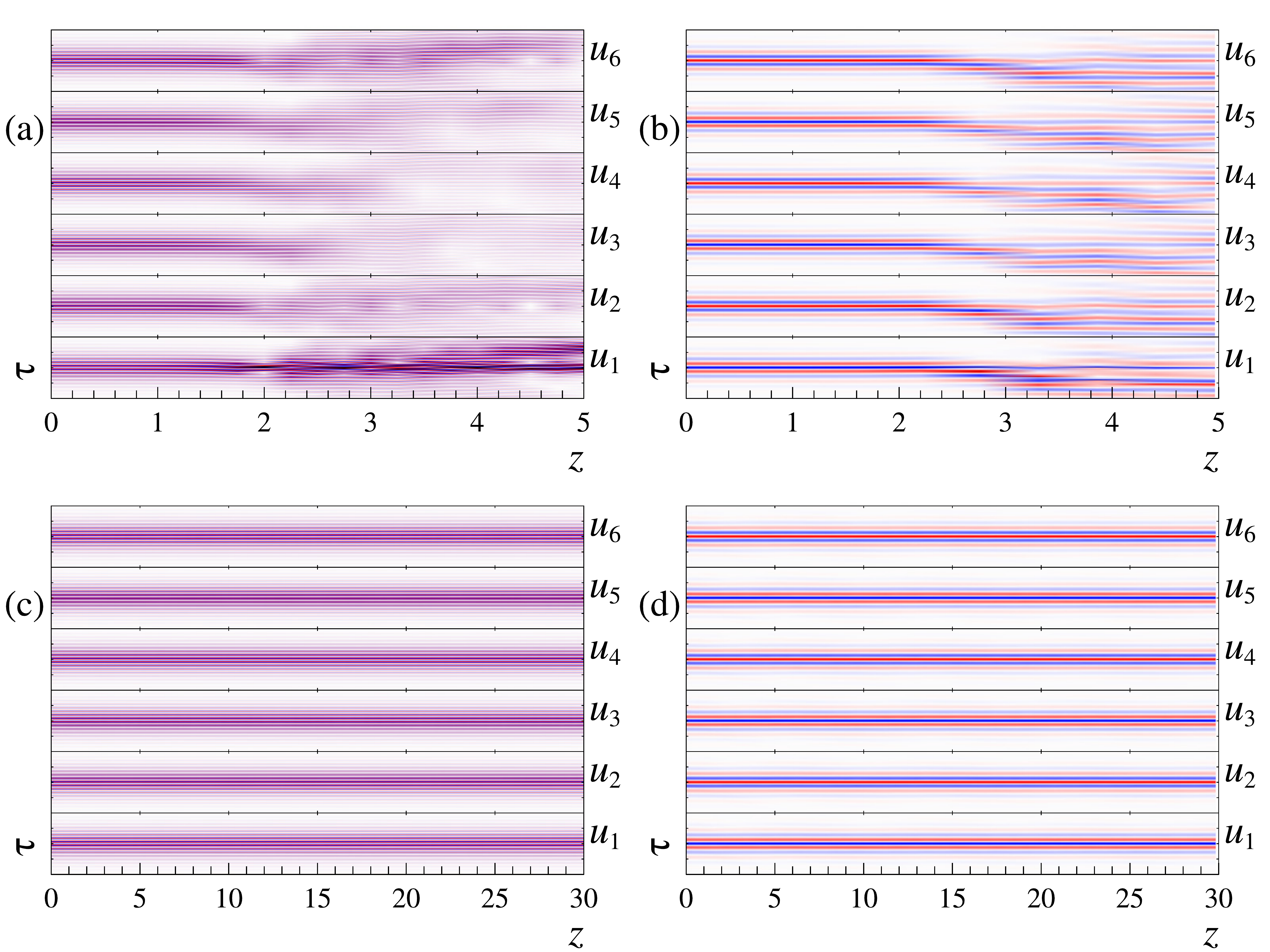}
\caption{
	Evolution of one of the components of the electric field strength $\text{Re}(u_n)$ of the found solution \eqref{eq:24} in an MCF of 6 cores at different values of $\delta $ and $\chi $: {\bf (a)} $\delta=0.05$, $\chi=0.002$, {\bf (b)} $\delta=0.3$, $\chi=0.2$, {\bf (c)} $\delta=0.05$, $\chi=0.008$, {\bf (d)} $\delta=0.3$, $\chi=0.3$. The initial noise level is $10^{-2}$.}\label{ris:ris2}
\end{figure*}

\section{Stability of out-of-phase spatio-temporal soliton}\label{sec:4}

Next, we turn to the question of stability of the found out-of-phase soliton \eqref{eq:24} with respect to spatial filamentation instability. Unfortunately, due to the complexity of the original equation \eqref{eq:16}, it is not possible to analyze analytically the stability of the solution found in MCF in the case of an inertialess Kerr nonlinearity ($f_R = 0$) and low-frequency dispersion ($\mu = 0 $).

First, let us establish a relation between the obtained solution \eqref{eq:24} and the out-of-phase envelope soliton solution, which was found earlier in the framework of the nonlinear Schr\"odinger equation \eqref{eq:17} \cite{Skobelev2019}. For this, we expand the first integral \eqref{eq:23b} in powers of $G$, since the transition to long quasi-monochromatic pulses corresponds to the case of small amplitudes $G \ll 1$. Keeping the terms of order $G^4$ in Eq.~\eqref{eq:23b}, we obtain a solution for the envelope $G(\xi)$, which corresponds to the Schr\"odinger soliton
\begin{multline}\label{eq:25}
	u_n^\text{sol}=(-1)^n\frac{\delta}{\omega_s}\frac{\sqrt{2(1+2\chi)} \rme^{\rmi\theta}}{\cosh(\delta\xi)} , \quad \delta\ll1 \\ \theta=\xi+\frac{2(1+2\chi)(1-\delta^2)z}{\omega_s}+\varphi_0.
\end{multline}
This solution coincides with the solution \eqref{eq:17}, obtained in the framework of NSE \cite{Skobelev2019}, if we set $\omega_s = 1$ and $\chi \ll 1$. The maximum amplitude exceeds the NSE soliton amplitude by factor $\sqrt{1 + 2 \chi}$ since we take into account in the equation \eqref{eq:19} that the group velocity dispersion is higher in the case of an out-of-phase mode \cite{Skryabin_2009}. Note that the velocity of a soliton does not depend on its amplitude and is equal to $1/v_s \simeq \omega_s^2$. Typical distributions of the field and spectral intensity of the soliton in one of the MCF cores at the small value of $\delta = 0.06$ are shown in the figure \ref{ris:ris1}{\bf (a, b)}.

As the amplitude increases, the soliton duration decreases and the processes associated with the dependence of the group velocity on the amplitude begin to play an increasing role, which is primarily reflected in the phase-modulated structure of the pulse \eqref{eq:23a}. Its amplitude dependence is then transformed into solitons of the generalized nonlinear Schr\"odinger equation, which have the following form, as is easy to obtain from \eqref{eq:23b} by expanding the polynomials and keeping the terms of the order of smallness $G^6$:
\begin{multline}\label{eq:26}
	u_n^\text{sol}=\frac{(-1)^n}{\omega_s\sqrt{1+\delta^2}}\frac{ 2\delta\sqrt{1+2\chi} \rme^{\rmi \theta}}{\sqrt{1+\sqrt{1+12\delta^2}\cosh(2\delta\xi)}} ,\\
		\quad \theta=\xi+2v_s\omega_sz +\frac32 \int |u_n^\text{sol}|^2 d\xi +\varphi_0.
\end{multline}
It follows from this expressions that the amplitude distribution of the soliton \eqref{eq:26} is close to the NSE soliton \eqref{eq:17}. However, a distinctive feature of the solution is the presence of a sufficiently strong frequency modulation in the laser pulse. Typical distributions of the field and spectral intensity at $\delta = 0.3$ are shown in the figure \ref{ris:ris1}{\bf (c, d)}.

In this case, the soliton velocity
\begin{equation}\label{eq:27}
	\frac1{v_s} = \frac{\omega_s^2(1+\delta^2)}{1+2\chi}
\end{equation}
depends on the amplitude ($\propto\delta^2$), which is a qualitative difference from the NSE solitons \eqref{eq:17}. Obviously, this fact can lead to the development of instability. Let us clarify this point. Let an out-of-phase spatial-temporal soliton be injected into the nonlinear medium. At this, the amplitudes in different cores are slightly different $\langle (\delta -\langle \delta\rangle)^2\rangle\neq 0$. Obviously, this will lead to a spread in the velocities of the solitons \eqref{eq:27} in different cores. Therefore, the found solution will be stable if the coupling length $2\pi/\chi$ is less than the dispersion length $2\omega_s/\delta^2$, i.e. for a sufficiently large coupling between neighboring cores
\begin{equation}\label{eq:28}
	\chi  > \chi_\text{cr}\equiv\frac{\pi \delta^2}{\omega_s} .
\end{equation}

To confirm the obtained estimate of the stability boundary \eqref{eq:28} of the found solution \eqref{eq:24}, let us turn to the results of numerical simulation. Figure \ref{ris:ris2} shows the evolution of one of the components of the electric field strength $\text{Re}(u_n)$ of a laser pulse with an initial distribution for $\omega_s = 1$ in an MCF of 6 cores for different values of the parameters $\delta$ and $\chi$. The initial noise level is $10^{-2}$.

Figure \ref{ris:ris2}{\bf (a)} shows the evolution of a wave packet at $\delta = 0.05$, which contains a large number of field oscillations. It can be seen that for the coupling coefficient $\chi = 0.002 < \chi_\text{cr}$ the solution \eqref{eq:24} is subject to filamentation instability and is destroyed as it propagates in the medium. However, for a larger coupling coefficient $\chi = 0.008> \chi_\text{cr}$ (Fig. \ref{ris:ris2}{\bf (c)}), the soliton solution propagates unchanged. Along with this, the figure \ref{ris:ris2}{\bf (b)} also shows the evolution of a wave packet at $\delta = 0.3$, which contains a pair of field oscillations. The solution is unstable for a small value of the coupling coefficient $0.2 <\chi_\text{cr}$, and become stable when the threshold $\chi = 0.3> \chi_\text{cr}$ is exceeded (Fig. \ref{ris:ris2}{\bf (d)}).

\begin{figure}[tp]
\centering
\includegraphics[width = \linewidth]{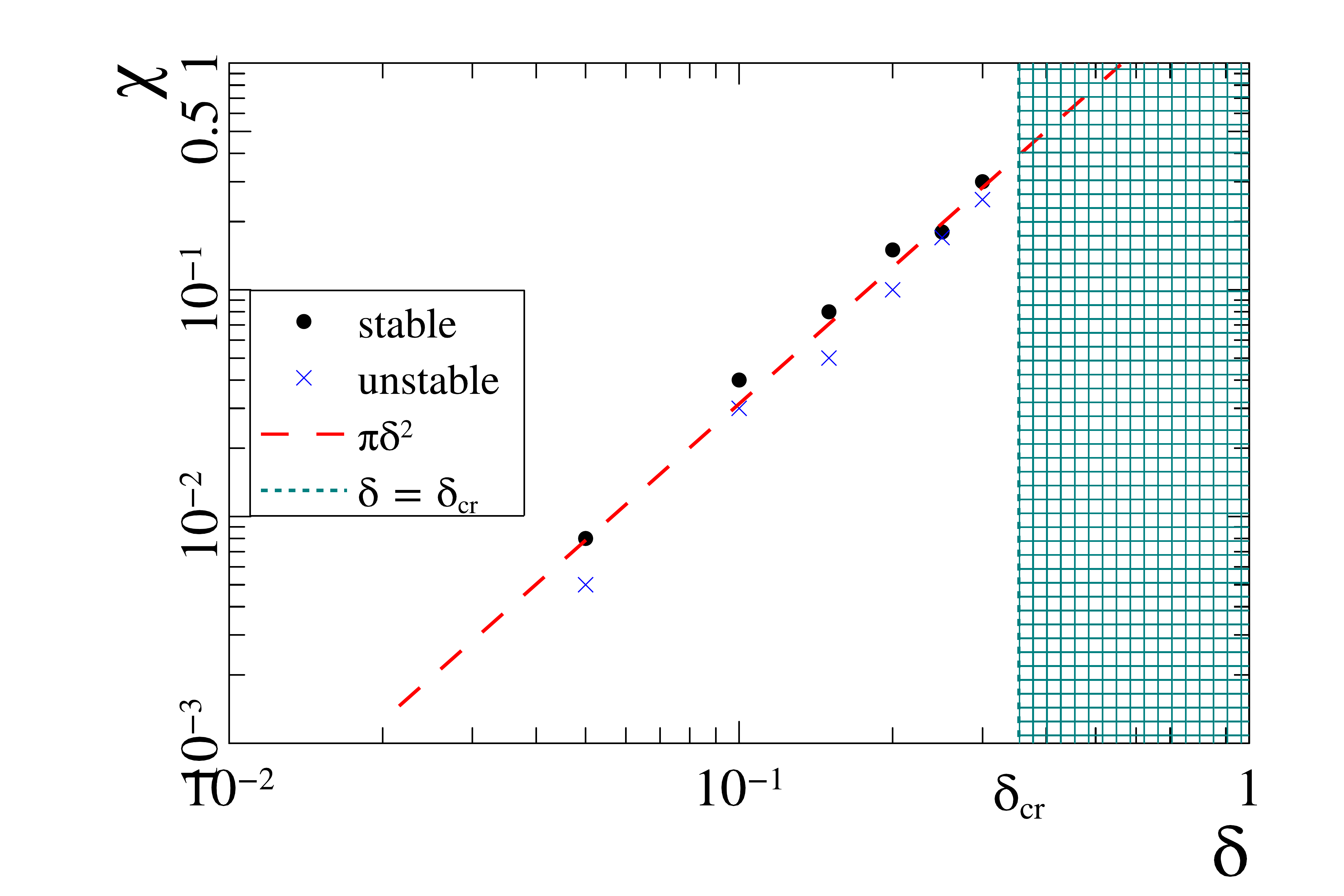}
\caption{
	Dependence of the threshold value of the coupling coefficient \eqref{eq:28} on the soliton parameter $\delta$ at $\omega_s = 1$. Dots show the parameters \eqref{eq:24}, which have shown their stability in numerical simulation, and the crosses correspond to an unstable propagation regime. The shaded area is where the soliton existence is impossible.}\label{ris:ris3}
\end{figure}

Figure \ref{ris:ris3} shows a map of the existence and stability of found solutions. As noted above, out-of-phase spatial-temporal soliton solutions exist for $\delta^2\leq 1/8$. In the figure, the red dashed line shows the stability boundary \eqref{eq:28}, dots and crosses show the results of numerical simulation. Thus, the found stability estimate is in good agreement with the results of numerical simulation.

\section{Laser pulse self-compression}\label{sec:5}

The studies carried out above have shown the existence of stable out-of-phase spatio-temporal soliton with a duration of up to one field oscillation in an MCF of $2N$ cores arranged in a ring. The found solution guarantees the coherent propagation of wave packets of unchanged shape in all cores of such an MCF. In this case, the total energy of the found nonlinear structure is $2N$ times the energy of a soliton in a single core.

In this section, we present the results of generalizing the well-known method of self-compression of laser pulses, based on multisoliton dynamics \cite{Agrawal_2019}, as applied already to MCFs. This will make it possible to take a significant step towards solving the problem of the formation of high-energy and short-duration laser pulses in systems built entirely in fiber.

A few words should be noted about the self-compression of laser pulses in the multisoliton dynamics regime in the case of single core \cite{Agrawal_2019}. Analysis of the NLS equation showed that in the case of a high-order soliton injected into the input of a nonlinear medium
\begin{equation}\label{eq:41}
	u=\frac{M\sqrt{2}\delta_0}{\cosh(\delta_0 \tau)}
\end{equation}
at the initial stage, a significant shortening of the laser pulse as a whole is observed. Subsequently, the wave packet decays into $M$ solitons with parameters $\delta_m=(2m-1)\delta_0$, where $m = 1 \ldots M$ is integer. Thus, a soliton is formed $ 2M-1 $ times shorter than the initial one with a fraction of the energy
\begin{equation}\label{eq:42}
	\eta=\frac{2M-1}{M^2} .
\end{equation}

\begin{figure}[tpb]
\centering
\includegraphics[width = \linewidth]{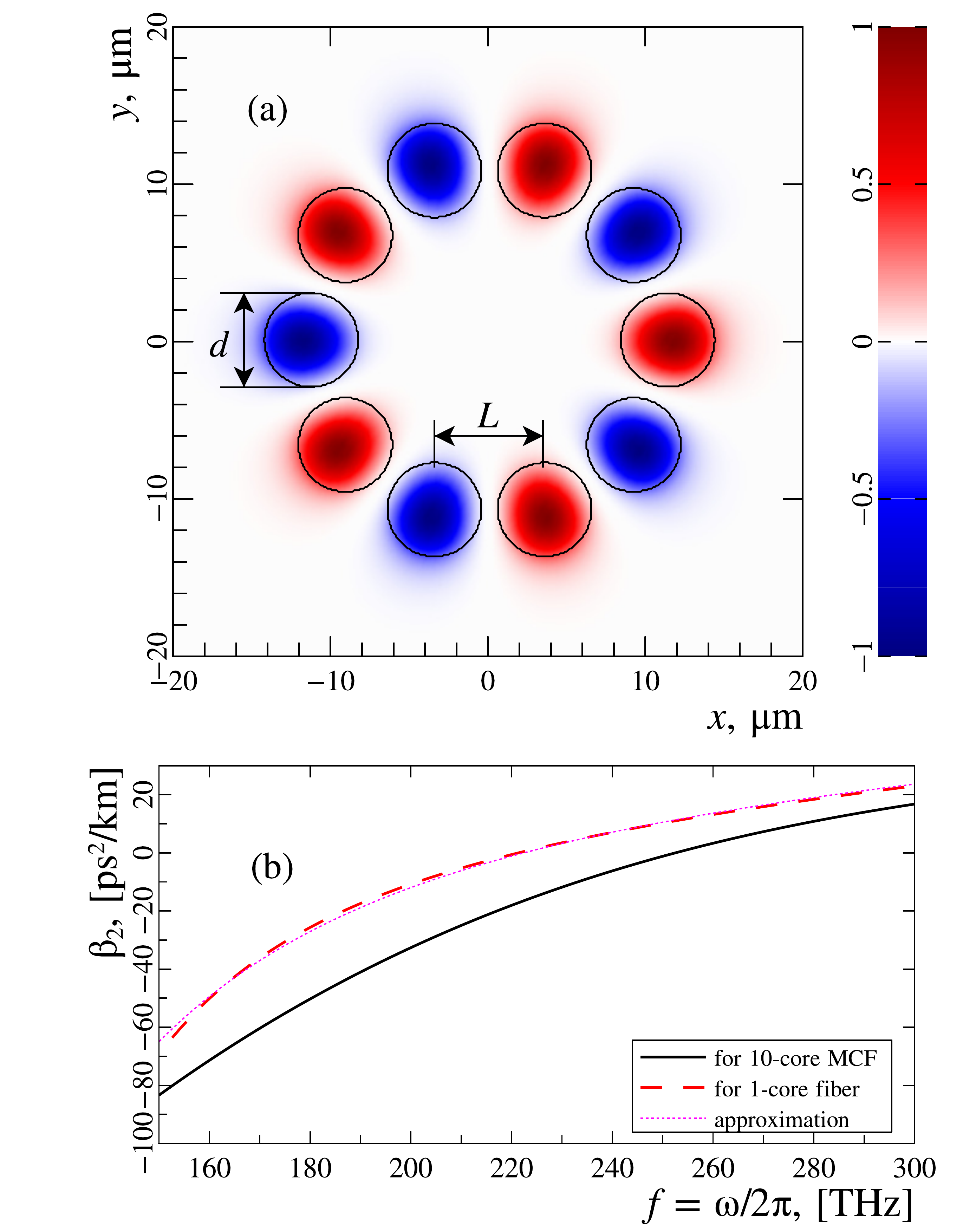}
\caption{
	{\bf (a)} Distribution of the $\mathcal{E}_x$ field at the wavelength $\lambda = 1550~\mu\text{m}$. The diameter of the cores is $d=6~\mu\text{m}$, the distance between the centers of the cores is $L = 7~\mu\text{m}$. {\bf (b)} Frequency dependence of the dispersion of the group velocity $\beta_2$: for 10-cores MCF (black solid), for a single core (red dash) and its approximation \eqref{eq:beta2}.}\label{ris:ris4}
\end{figure}

Next, we turn to a numerical analysis of the self-compression of a laser pulse in an MCF consisting of 10 cores, arranged in a ring (Fig. \ref{ris:ris4}{\bf (a)}). At the MCF input, we inject a laser pulse with an initial duration of 100 fs, whose amplitude is three times greater than the amplitude of the found out-of-phase spatio-temporal soliton. It is expected that three pulses should be formed in the process of nonlinear dynamics. In this case, the duration of the high-intensity soliton should be approximately five times shorter than the initial duration.

Consider an MCF with a silica cladding and cores doped with GeO$_2$ 6\%, which gives a refractive index difference of $\Delta n \sim 0.009$ compared to the cladding one. The diameters of the cores $d$ and the distances between them $L$ were chosen to satisfy the stability condition for the out-of-phase spatio-temporal soluton \eqref{eq:28}. In other words, the coupling length $L_b=2\pi/\chi$ must be less than the dispersion one $L_\text{dis}=\tau_\text{min}^2/|\beta_2|$, estimated for the soliton with the shortest duration (about $15 \ldots 20$~fs).

The use of a finite-element code allows us to find for such a configuration both the distribution of the field of the out-of-phase mode, and the propagation constant $K_\pm(\omega)$, as a function of frequency, and calculate the dispersion of the group velocity $\beta_2=\partial^2 K_\pm/\partial\omega^2$. Figure \ref{ris:ris4}{\bf (a)} shows an example of the distribution of the field $\mathcal{E}_x$ at the wavelength $\lambda = 1550~\mu\text{m}$. In the figure \ref{ris:ris4}{\bf (b)}, the black line shows the calculated $\beta_2(\omega)$ for $d = 6~\mu\text{m}$ and $L = 7~\mu\text{m}$. For comparison, the red dashed line in the figure shows the dependence of $\beta_2(\omega)$ in the case of single core. It can be seen that the dispersion of the group velocity $\beta_2$ for the out-of-phase mode in MCF is greater in absolute value than in the case of a single core, which is consistent with our earlier conclusion \eqref{eq:19}. Along with this, the dotted magenta curve shows the approximation \eqref{eq:beta2} of the group velocity dispersion for one core, which fits well on the red dashed line.

\begin{figure}[htpb]
\centering
\includegraphics[width = \linewidth]{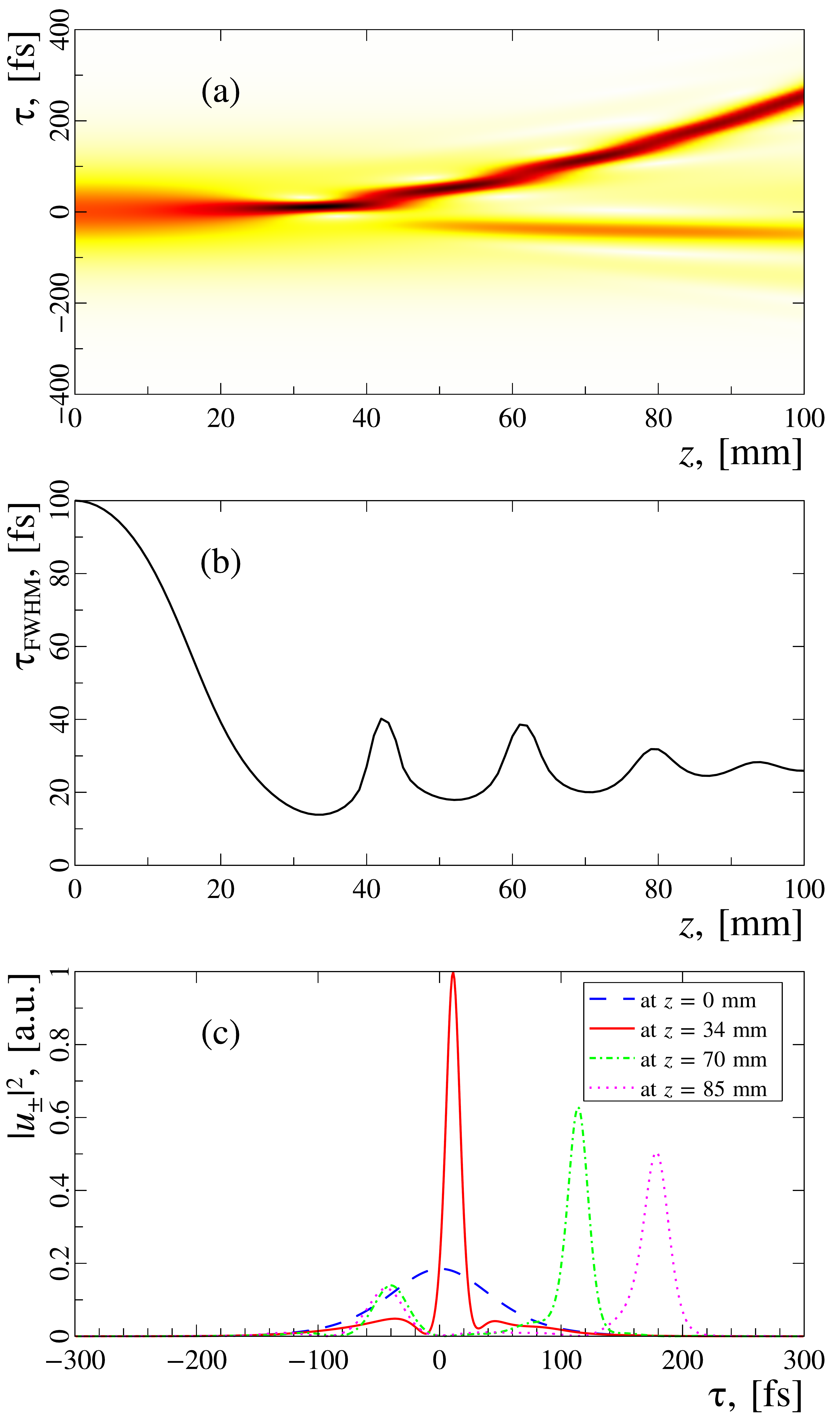}
\caption{
	{\bf (a)} Dynamics of the field envelope $|{u}_\pm(z,\tau)|$. {\bf (b)} Dependence of the wave packet duration on the evolutionary variable $z$. {\bf (c)} Intensity distribution of the wave packet $|{u}_\pm(\tau)|^2$ for different values of $z$ coordinate.}\label{ris:ris5}
\end{figure}

To further simplify the numerical calculation for the purpose of a detailed analysis of the nonlinear dynamics of a laser pulse, we use explicitly the stability of the out-of-phase solitary solution. We represent the electric field of a laser pulse in the form
\begin{equation} \label{Ew}
\vec{E}_\omega({\vec{r}}, z) = \vec{F}_\pm(\vec{r},\omega)\, g_m,
\end{equation}
where $\vec{E}_\omega(\vec{r},z) = \hat{F}[\vec{E}(\vec{r},z,\tau)]$ is the spectrum of the field laser pulse, $\hat{F}$ is the Fourier transform in the $\tau$ coordinate, $\vec{F}_\pm({\vec{r}}, \omega)$ is the spatial distribution of the mode. The dynamics of the laser pulse spectrum envelope $g_\pm = \hat{F}[{u}_\pm]$ will be described by a unidirectional wave equation taking into account the exactly found dispersion shown in Fig. \ref{ris:ris4}{\bf (b)}, Kerr and Raman nonlinearities, nonlinear dispersion
\begin{multline}\label{Eq}
\frac{\partial g_\pm}{\partial z} - \rmi K_\pm g_\pm  =  \rmi \gamma  \hat{F}\Bigg[\bigg(1+\frac{\rmi}{\omega_0} \frac{\partial}{\partial\tau}\bigg) \bigg((1-f_R) |{u}_\pm|^2 {u}_\pm +\\+
f_R {u}_\pm \int_{-\infty}^{\tau}h_R(\xi)|{u}_\pm(\tau-\xi)|^2d\xi \bigg) \Bigg].
\end{multline}
where $\omega_0$ is the center frequency of the laser pulse, $f_R$ and $h_R$ are determined in the same way \eqref{eq:15}. The main difference of this equation from the Eq.~\eqref{eq:16} considered above is the exact account of the dispersion of the medium --- the dependence of the wave number $K_\pm(\omega)$ on the frequency was calculated using a finite-element code.

Figure \ref{ris:ris5} shows the results of numerical simulation of the dynamics of a wave packet in an MCF of ten cores with $d = 6~\mu\text{m}$ and $L = 7~\mu\text{m}$, with a length of ten cm. A laser pulse with an energy of 40 nJ and a duration of 100 fs at a wavelength of $\lambda=1550~\mu\text{m}$ was injected into the MCF. The spatial distribution of the wave field is determined by the out-of-phase mode. For a given MCF, the coupling length is $L_b=2\pi/\chi=1.2$~mm. The effective area of the out-of-phase supermode at this wavelength is $300~\mu\text{m}^2$, the nonlinear coefficient $\gamma=0.3 1/(\text{W}\cdotp\text{km})$.

Figure \ref{ris:ris5}{\bf (a)} shows the evolution of the envelope of the wave packet $|{u}_\pm|$. Figure \ref{ris:ris5}{\bf (b)} shows the dependence of the wave packet duration on the evolutionary variable $z$. It is seen that as the evolution of the laser pulse in a nonlinear medium, an adiabatic decrease in the duration of the wave packet takes place. The wave packet is shortened as much as possible and reached duration of $\tau_p = 14$~fs at a length of $z = 34$~mm. Figure \ref{ris:ris5}{\bf(c)} shows the intensity distributions of the wave packet. The initial distribution is shown by the blue dashed line, the red line shows the distribution of the wave packet at the length $z = 34$~mm. In this figure, the value of the intensity of the laser pulse is normalized to the maximum value. Note that the dispersion length of the compressed laser pulse is 5.2 mm, which exceeds the coupling length $L_b = 1.2$~mm. The energy in the compressed pulse is more than 38~nJ.

Figure \ref{ris:ris5}{\bf (a)} shows that the laser pulse is subsequently split into three wave structures. The most intense of them forms a soliton with a duration 5 times shorter than the initial one, which is in good agreement with the above theoretical picture. In this case, the wave packet with the maximum amplitude (with the shortest duration) is rather quickly separated from the remaining structures, since their group velocities differ significantly. This separation of pulses is strengthens by the shifting of the center frequency due to Raman nonlinearity. From the figures \ref{ris:ris5}{\bf (a,b)}, it can be seen that already at distances of $z \gtrsim 70$~mm, the high-intensity soliton is well isolated from the remaining two structures and further propagates stably, without any changes.

\section{Conclusion}\label{sec:9}

In this work, a {\it basic equation} for the analysis of key self-action features of wave packets with an arbitrary durations in the MCF is derived. This equation describes the evolution of the wave field in the MFC without scale division into slow envelope and high-frequency carrier. A new class of stable out-of-phase spatio-temporal few-cycle soliton in the MCF consisting of cores arranged in a ring is found and analyzed. The stability condition of the obtained solutions is determined. These nonlinear structures represent an extension of solutions found by us earlier in the framework of NSE to few-cycle regime. \cite{Skobelev2019} is demonstrated. This allows them to be considered as {\it elementary} wave field structures that play the same fundamental role in the nonlinear dynamics of wave fields as NSE solitons applied to a single fiber.

As an example of the such solitons use, we considered the problem of their self-compression in the process of multisoliton dynamics to effectively shorten laser pulses to a duration of several optical cycles in the MCF. In this paper, we consider the case that most closely matches the realizable experimental situation of compression of a laser pulse with an initial duration of 100 fs and an energy of 40 nJ at a wavelength of $1550~\mu\text{m}$, propagating in an MCF of ten cores arranged in a ring. As a result, as shown by numerical calculations, a laser pulse with a duration of 14~fs (slightly less than 3 field periods) with an energy of 38 nJ will be formed at the output of an MCF 34 mm long.

This work was supported by the Center of Excellence ``Center of Photonics'' funded by the Ministry of Science and Higher Education of the Russian Federation, contract No. 075-15-2020-906.

\end{document}